\documentclass[12pt,a4paper]{article}

\usepackage{graphics,graphicx,mhchem,placeins,afterpage,makecell,tablefootnote,multicol,multirow,caption,textcomp,lineno,tabularx,subfigure,booktabs,courier,collref,setspace,float,xcolor,amsmath,amssymb} 
\usepackage[utf8]{inputenc}
\usepackage[english]{babel}
\usepackage[backend=bibtex,style=authoryear,bibstyle=authoryear,maxcitenames=2,maxbibnames=7,useprefix=true]{biblatex}

\DeclareNameAlias{sortname}{last-first}

\begin{document}

\begin{center}
  {\Large{Detection of CH$_3$C$_3$N in Titan's
      Atmosphere}}
\end{center}

\begin{center}
Authors: Alexander E. Thelen$^{a,b,}$\footnote{Corresponding
  author. (A. E. Thelen) Email address: alexander.e.thelen@nasa.gov},
M. A. Cordiner$^{a,c}$, C. A. Nixon$^a$, V. Vuitton$^d$,
Z. Kisiel$^e$, S. B. Charnley$^a$, M. Y. Palmer$^f$, N. A. Teanby$^g$,
P. G. J. Irwin$^h$

{\footnotesize{$^a$Solar System Exploration Division, NASA Goddard Space
  Flight Center, Greenbelt, MD 20771, USA, $^b$Universities Space Research Association, Columbia, MD
  21046, USA, $^c$Department of Physics, Catholic University of America, Washington,
  DC 20064, USA, $^d$Institut de Plan{\'e}tologie et d'Astrophysique de
  Grenoble, Universit{\'e} Grenoble Alpes, CNRS, Grenoble 38000,
  France, $^e$Institute of Physics, Polish Academy of Sciences,
  Al. Lotnik{\'o}w 32/46, 02-668 Warszawa, Poland, $^f$Lunar and Planetary Laboratory, University of Arizona,
  Tucson, AZ 85721, USA, $^g$School of Earth Sciences, University of Bristol, Bristol
  BS8 1RJ, UK, $^h$Atmospheric, Oceanic and Planetary Physics, Clarendon Laboratory,
  University of Oxford, Oxford, OX1 3PU, UK}}

\end{center}

\begin{center}
  (Accepted October 15, 2020 -- Astrophysical Journal Letters)
\end{center}

\begin{abstract}
Titan harbors a dense, organic-rich atmosphere primarily composed of
N$_2$ and CH$_4$, with lesser amounts of hydrocarbons and
nitrogen-bearing species. As a result of high sensitivity observations by the Atacama Large
Millimeter/submillimeter Array (ALMA) in Band 6 ($\sim$230--272 GHz), we obtained the first
spectroscopic detection of
CH$_3$C$_3$N (methylcyanoacetylene or cyanopropyne) in
Titan's atmosphere through the observation of seven transitions in the
$J = 64\rightarrow63$ and $J = 62\rightarrow61$ rotational bands. The presence of
CH$_3$C$_3$N on Titan was suggested by the Cassini Ion and Neutral Mass
Spectrometer detection of its
protonated form: C$_4$H$_3$NH$^+$, but the atmospheric abundance of the associated (deprotonated)
neutral product is not well constrained due to the lack of appropriate
laboratory reaction data. Here, we derive the column density of CH$_3$C$_3$N to be
(3.8--5.7)$\times10^{12}$ cm$^{-2}$ based on radiative transfer models
sensitive to altitudes above 400 km Titan's middle
atmosphere. When compared
with laboratory and photochemical model results, the detection of
methylcyanoacetylene provides important
constraints for the determination of the associated production
pathways (such as those involving CN, CCN, and hydrocarbons), and reaction rate
coefficients. These results also further demonstrate the importance of
ALMA and (sub)millimeter spectroscopy for future investigations of
Titan's organic inventory and atmospheric chemistry, as CH$_3$C$_3$N
marks the heaviest polar molecule detected spectroscopically in Titan's
atmosphere to date.
\end{abstract}

\section{Introduction} \label{sec:intro}
\renewcommand{\thefootnote}{\arabic{footnote}}
\setcounter{footnote}{0}
The atmosphere of Titan, the largest moon of Saturn, is primarily
composed of N$_2$ ($\sim$95--98$\%$) and CH$_4$
($\sim$1--5$\%$). A plethora of trace organic compounds makes up the remaining
atmospheric composition, which are formed through the photodissociation of
nitrogen and methane, and interactions with the Saturnian magnetosphere or galactic cosmic rays
(GCR) \parencite{loison_15, vuitton_19}. The
formation of complex atmospheric species -- such as nitriles
(C$_X$H$_Y$[CN]$_Z$) -- in Titan's upper atmosphere, their condensation
and accumulation in the stratospheric haze layer, and
participation in the methane-based meteorological cycle, are important processes that influence
not only Titan's
global climate but also the connection
between the atmosphere and the organic regolith and hydrocarbon lakes. In
addition to increasing our understanding of Titan's atmospheric and
surface properties, knowledge of Titan's atmospheric photochemistry
and the extent of its molecular inventory are important for assessing
Titan's potential for habitability \parencite{horst_12, palmer_17}.

While numerous heavy ion species were detected with the Ion and
Neutral Mass Spectrometer (INMS) and Cassini Plasma Spectrometer instruments onboard the Cassini
spacecraft at altitudes $\sim$1000--1200 km, these measurements did not fully resolve the identities of
many large species -- particularly those with a mass-to-charge ratio ($m/z$) $>60$. Among these, the detection of ions with $m/z = 66$,
attributed to C$_4$H$_3$NH$^+$ \parencite{vuitton_07}, presented the case for multiple
associated neutral species: CH$_3$C$_3$N (methylcyanoacetylene or cyanopropyne,
hereafter the former) or H$_2$CCCHCN
  (cyanoallene). Laboratory
experiments predict the formation of both C$_4$H$_3$N isomers in Titan's
atmosphere through crossed molecular beam \parencite{huang_99,
  balucani_00b} and plasma discharge \parencite{thompson_91, coll_99,
  molina-cuberos_02} experiments under Titan-like
(N$_2$/CH$_4$-rich) conditions. However, while
both methylcyanoacetylene and cyanoallene have been detected
previously in the interstellar medium toward the Sgr B2 high-mass
star-forming region and in the
nearby molecular cloud TMC-1
\parencite{broten_84, lovas_06, belloche_13}, the C$_4$H$_3$N isomers have
yet to be detected in the atmosphere of Titan. 

The advent of the Atacama Large Millimeter/submillimeter Array
(ALMA) in the past decade has enabled the exploration of Titan's atmospheric
composition and dynamics to an unprecedented degree from the ground,
allowing for follow-up studies on the distribution of trace molecular
species by the \textit{Voyager-1} and Cassini-Huygens missions.
Comprised of many 12 m antennas spatially separated by up to 16
km and access to frequencies ranging from $\sim$84--950 GHz ($\sim$3.5--0.3 mm),
ALMA has enabled the detection of new molecular species
\parencite{cordiner_15, palmer_17, nixon_20}, isotopes 
\parencite{serigano_16, molter_16, thelen_19b, iino_20}, and
vibrationally excited transitions \parencite{cordiner_18, kisiel_20} in
Titan's atmosphere. The spectral and spatial resolution
capabilities of ALMA have also provided the means by which to map the
distribution and dynamics of many nitrogen-bearing molecules \parencite{cordiner_14,
  lai_17, thelen_19, cordiner_19, lellouch_19}, allowing for the study of
atmospheric variations throughout Titan's long (29.5 yr) seasonal cycle after
the end of the Cassini mission in 2017.

Here, we detail the first results of deep ALMA Cycle 8 observations of Titan during
November and December, 2019. The
high sensitivity of these data have allowed for the spectroscopic 
detection of two CH$_3$C$_3$N bands for the first time in Titan's
atmosphere (or indeed the atmosphere of any Solar System body).

\section{Observations} \label{sec:obs}
Titan was observed across multiple execution blocks between
November 14 and December 16, 2019, under ALMA Project Code
$\#$2019.1.00783.S. Integrations on Titan were distributed across three
Science Goals (SG). SG1, which targeted the CO $J =
2\rightarrow1$ transition at 230.538 GHz to retrieve Titan's
disk-averaged temperature profile, was observed on November 14, 2019
for 11.59 min in ALMA configuration C43-3
    (maximum baselines of 500 m)
with 44 antennas. SG2 and SG3
covered multiple nitrile species, their C- and N-isotopes
(e.g. H$^{13}$CCCN, HCCC$^{15}$N), and potential isocyanide
species. Observations for these two Science Goals required seven executions between
November 25 to December 16, 2019, with 43--45 antennas
in ALMA configurations C43-1 and C43-2
    (maximum baselines ranging from $\sim$160--314 m); the
cumulative integration time on Titan was 81.65 min for SG3 and
175.4 min for SG2, which required the highest spectral
sensitivity ($\sim$1 mJy). Spectra from all three Science Goals were
analyzed for the detection and subsequent radiative transfer
modeling of CO and CH$_3$C$_3$N transitions.

ALMA visibility data were calibrated with version 5.6.1-8 of the Common Astronomy Software Applications
(CASA) pipeline using the scripts provided by the Joint ALMA
Observatory (JAO). In addition to Titan, the quasars J1924-2914,
J1911-2006, and J2056-4714 were also observed for the purposes of
flux, bandpass, and phase calibrations. Subsequent executions of the
pipeline calibrations were completed after modifying
the JAO scripts to implement a variety
of bandpass smoothing intervals to improve the spectral
root mean square (RMS) noise (particularly in SG2, with the longest
total integration time) without significantly degrading the bandpass
shape. See Appendix \ref{sec:app} for the results and discussion of the effects of bandpass
smoothing on these ALMA observations.

\begin{figure}
  \centering
  \includegraphics[scale=0.85]{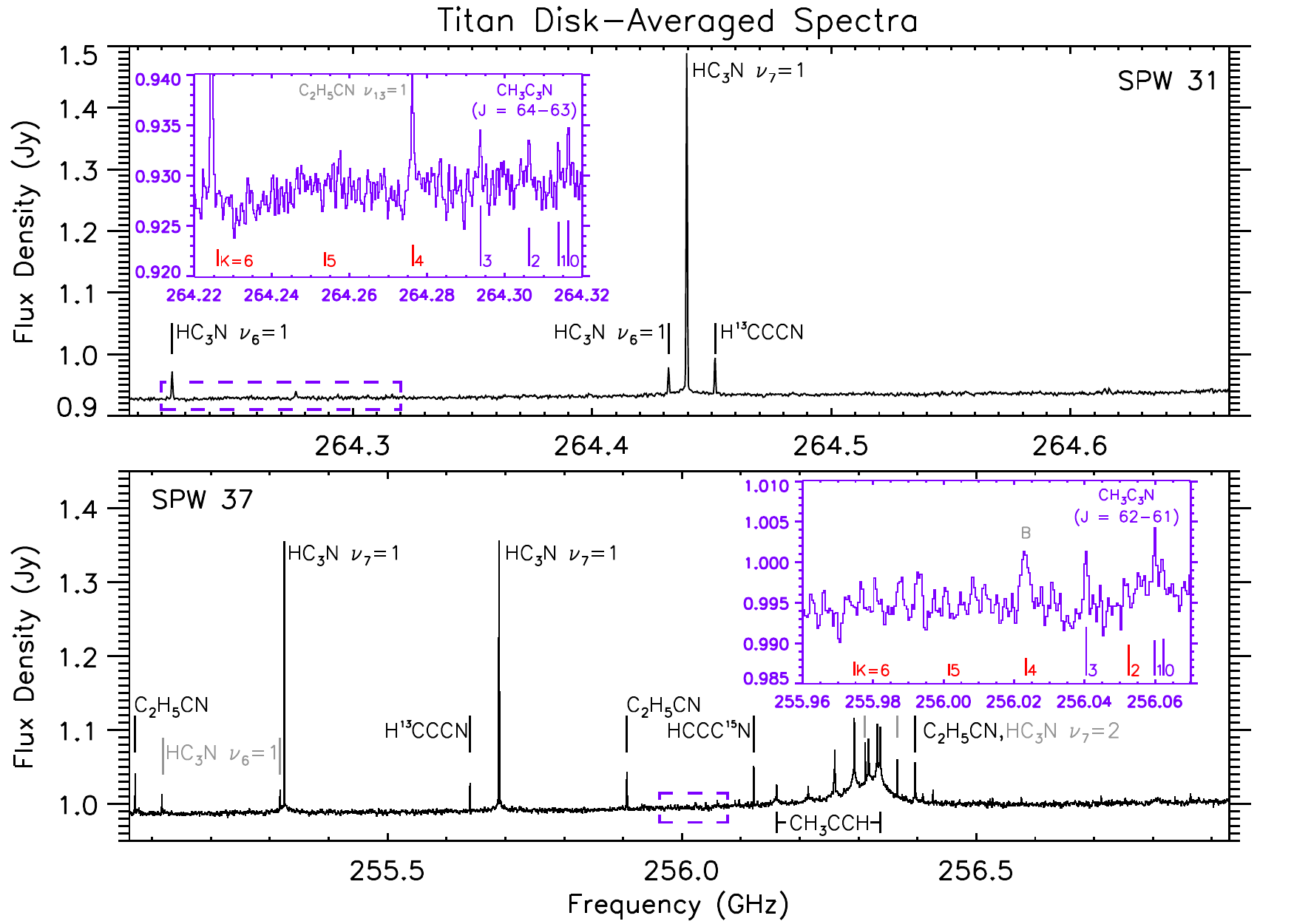}
  \caption{Disk-averaged spectra of Titan from SG2 and SG3 spectral
  windows 31 (top) and 37 (bottom), respectively. Strong spectral lines of
  various molecular species are marked with black or gray lines; spectral
  line parameters are detailed in Table 1. Additional unlabeled transitions of
  C$_2$H$_5$CN and C$_2$H$_3$CN are present. Insets in purple show the
  detections of the CH$_3$C$_3$N $K = $ 0--3 lines in the $J = 64\rightarrow63$ and $J =
  62\rightarrow61$ bands. Both detected transitions (purple) and
  undetected or blended transitions (red) are marked, with marker
  heights proportional to the line intensities (calculated at 160 K). An additional, blended
feature (B) is shown in the inset of SPW 37 at $\sim256.024$ GHz, most likely a combination of
C$_3$H$_8$, C$_2$H$_5$CN, and CH$_3$C$_3$N.}
  \label{fig:spec}
\end{figure}

The CASA \texttt{tclean} procedure was performed on the resulting
calibrated visibility measurement sets to deconvolve the complex
interferometric point-spread function (PSF) and reconstruct the brightness
distribution of Titan in standard spatial coordinates. The
\texttt{H{\"o}gbom} algorithm was used during deconvolution with
``natural'' weighting applied to prioritize the image signal-to-noise
ratio (SNR). The
\texttt{tclean} image sizes and pixel scales were set to
$270\times270$ $0.16''$-pixels for SG1, $224\times224$ $0.17''$-pixels for SG2, and
$210\times210$ $0.18''$-pixels for SG3, so as to sufficiently sample
the ALMA PSF ($\sim$5 pixels across the full width at half
maximum). Images produced through concatenated integrations
(SG2, SG3) were set to use a common synthesized beam shape (the ALMA PSF), and were
corrected for the ALMA primary beam. The resulting images had beam
sizes equal to $1.276''\times0.933''$ for SG1,
$1.470''\times1.067''$ for SG2, and $1.553''\times1.142''$ for SG3. As
Titan's angular diameter (with its extended atmosphere up to 1200 km)
was between $0.954''$--$0.985''$ during these observations, we were unable to
investigate potential spatial variation in the atmospheric
distribution of CH$_3$C$_3$N from these images. The resulting disk-averaged spectra of
spectral windows (SPW) 31 (SG2) and 37 (SG3) are shown in Figure
\ref{fig:spec}, including the detections of the CH$_3$C$_3$N  $J =
64\rightarrow63$ (SNR =
    3.42--4.27$\sigma$) and $J = 62\rightarrow61$ bands (SNR =
    2.37--4.58$\sigma$), respectively, in the panel
insets. The inherent channel spacing of the SG2 and SG3 spectral
windows were 244 and 488 kHz, respectively, resulting
in spectral resolutions of 488 and 976 kHz after Hanning smoothing by
the correlator. A
number of additional transitions from other molecular
species were detected in these two spectral windows, which are
detailed in Table 1. While the CH$_3$C$_3$N $J = 62\rightarrow61$, $K=2$
transition at 256.053 GHz was not detected above the noise in SG3 SPW 37, seven
other CH$_3$C$_3$N transitions were detected between both spectral
windows (all but one of which were detected at
    greater than $3\sigma$ confidence level -- see Appendix \ref{sec:app}). 

    \begin{table}
      \begin{center}
        \caption[]{Spectral Transitions}
        \begin{tabular}{lccrcc}
          \toprule
          \textbf{Species} & \textbf{Rest Freq.} & \textbf{Transition$^a$} & \textbf{$E''$} & \textbf{SG} & \textbf{Spw} \\
                           & \textbf{(GHz)} & & \textbf{(K)} & & \\
          \midrule
          \midrule
          CO & 230.538 & $2\rightarrow1$ & 17 & 1 & 29 \\
          
          HC$_3$N & 255.116 & $28\rightarrow27$, $\nu_6=1f$& 895 & 3 & 37 \\
          HC$_3$N & 255.317 & $28\rightarrow27$, $\nu_6=1e$ & 895 & 3 & 37 \\
          HC$_3$N & 264.224 & $29\rightarrow28$, $\nu_6=1e$ & 908 & 2 & 31 \\
          HC$_3$N & 264.431 & $29\rightarrow28$, $\nu_6=1f$ & 908 & 2 & 31 \\
          
          HC$_3$N & 255.324 & $28\rightarrow27$, $\nu_7=1f$ & 499 & 3 & 37 \\
          HC$_3$N & 255.689 & $28\rightarrow27$, $\nu_7=1e$ & 499 & 3 & 37 \\
          HC$_3$N & 264.439 & $29\rightarrow28$, $\nu_7=1e$ & 511 & 2 & 31 \\
          
          HC$_3$N & 256.311 & $28\rightarrow27$, $\nu_7=2f$ & 823 & 3 & 37 \\
          HC$_3$N & 256.365 & $28\rightarrow27$, $\nu_7=2e$ & 823 & 3 & 37 \\
          
          H$^{13}$CCCN & 255.639 & $29\rightarrow28$ & 184 & 3 & 37 \\
          H$^{13}$CCCN & 264.451 & $30\rightarrow29$ & 197 & 2 & 31 \\
          
          HCCC$^{15}$N & 256.121 & $29\rightarrow28$ & 184 & 3 & 37 \\
          
          C$_2$H$_5$CN & 255.071 & $28_{2,26}\rightarrow27_{2,25}$ & 182 & 3 & 37 \\
          C$_2$H$_5$CN & 255.906 & $28_{3,25}\rightarrow27_{3,24}$ & 186 & 3 & 37 \\
          C$_2$H$_5$CN & 256.396 & $29_{1,28}\rightarrow28_{1,27}$ & 189 & 3 & 37 \\
          C$_2$H$_5$CN & 264.276 & $29_{4,26}\rightarrow28_{4,25}$, $\nu_{13} = 1$ & 502 & 2 & 31 \\
          
          CH$_3$CCH & 256.097 & $15_7\rightarrow14_7$ & 452 & 3 & 37 \\
          CH$_3$CCH & 256.161 & $15_6\rightarrow14_6$ & 358 & 3 & 37 \\
          CH$_3$CCH & 256.214 & $15_5\rightarrow14_5$ & 279 & 3 & 37 \\
          CH$_3$CCH & 256.258 & $15_4\rightarrow14_4$ & 214 & 3 & 37 \\
          CH$_3$CCH & 256.293 & $15_3\rightarrow14_3$ & 163 & 3 & 37 \\
          CH$_3$CCH & 256.317 & $15_2\rightarrow14_2$ & 127 & 3 & 37 \\
          CH$_3$CCH & 256.332 & $15_1\rightarrow14_1$ & 106 & 3 & 37 \\
          CH$_3$CCH & 256.337 & $15_0\rightarrow14_0$ & 98 & 3 & 37 \\
          
          \\

          \textcolor{red}{CH$_3$C$_3$N} & 255.975 & $62_6\rightarrow61_6$ & 656 & 3 & 37 \\
          \textcolor{red}{CH$_3$C$_3$N} & 256.001 & $62_5\rightarrow61_5$ & 574 & 3 & 37 \\
          \textcolor{red}{CH$_3$C$_3$N} & 256.023 & $62_4\rightarrow61_4$ & 507 & 3 & 37 \\
          CH$_3$C$_3$N & 256.040 & $62_3\rightarrow61_3$ & 455 & 3 & 37 \\
          \textcolor{red}{CH$_3$C$_3$N} & 256.053 & $62_2\rightarrow61_2$ & 417 & 3 & 37 \\
          CH$_3$C$_3$N & 256.060 & $62_1\rightarrow61_1$ & 395 & 3 & 37 \\
          CH$_3$C$_3$N & 256.062 & $62_0\rightarrow61_0$ & 387 & 3 & 37 \\
          \textcolor{red}{CH$_3$C$_3$N} & 264.226 & $64_6\rightarrow63_6$ & 682 & 2 & 31 \\
          \textcolor{red}{CH$_3$C$_3$N} & 264.254 & $64_5\rightarrow63_5$ & 599 & 2 & 31 \\
          \textcolor{red}{CH$_3$C$_3$N} & 264.276 & $64_4\rightarrow63_4$ & 532 & 2 & 31 \\
          CH$_3$C$_3$N & 264.294 & $64_3\rightarrow63_3$ & 480 & 2 & 31 \\
          CH$_3$C$_3$N & 264.306 & $64_2\rightarrow63_2$ & 442 & 2 & 31 \\
          CH$_3$C$_3$N & 264.314 & $64_1\rightarrow63_1$ & 420 & 2 & 31 \\
          CH$_3$C$_3$N & 264.316 & $64_0\rightarrow63_0$ & 412 & 2 & 31 \\
          \bottomrule
          \end{tabular}
        \end{center}
        {\small{{\bf{Notes:}} Rows in red denote undetected (often higher energy)
          CH$_3$C$_3$N transitions. CH$_3$C$_3$N line positions were taken
          from the CDMS catalogue. $^a$Rotational transitions are written
          as $J'' \rightarrow J'$, $J''_{K''} \rightarrow
          J'_{K'}$, or $J''_{K_a'', K_c''} \rightarrow J'_{K_a',
          K_c'}$.}}
  \end{table}

\section{Radiative Transfer Modeling $\&$ Results} \label{sec:rad}
Disk-averaged data were extracted from spectral image cubes
using a circular mask that encompassed pixels with $\geq90\%$
of Titan's continuum flux \parencite{lai_17,thelen_19b,nixon_20}. Variations in Titan's distance and relative
velocity between integrations were accounted for in the previous calibration and imaging
steps. We used 36 line-of-sight emission angles to properly
characterize Titan's disk-averaged radiance from the surface to 1200
km \parencite{teanby_13, thelen_18, thelen_19}, and applied small
multiplicative factors to the spectra to resolve differences between
the data and synthetic spectra in continuum regions (scaling factors
on the order 1.15; see \cite{thelen_18}). We employed the
Non-linear optimal Estimator for Multi-variatE
spectral analySIS (NEMESIS) radiative transfer package
\parencite{irwin_08} to model
Titan's atmospheric emission and retrieve vertical temperature and
volume mixing ratio profiles, as has been used previously for Cassini Composite
Infrared Spectrometer and ALMA observations of Titan
(see, for example, \cite{nixon_10}, \cite{teanby_10a}). The
NEMESIS atmospheric model parameterization we used
follows the prescription of previous studies of Titan with ALMA
\parencite{thelen_18, thelen_19b, thelen_19}. Spectral line parameters from the Cologne Database for Molecular
Spectroscopy (CDMS; \cite{muller_01, muller_05, endres_16}) were
used for models of CH$_3$C$_3$N
(\cite{moises_82,bester_84,bester_85}; with purely $K$-dependent
line parameters taken from CH$_3$CN,
\cite{anttila_93}). The excited
state C$_2$H$_5$CN lines not yet available in CDMS
(e.g. Fig. \ref{fig:spec}, top inset) were taken from
\textcite{kisiel_20}. We assumed values
    for the CH$_3$C$_3$N Lorentzian broadening half-width ($\Gamma$) = 0.115 cm$^{-1}$
bar$^{-1}$, and temperature dependence exponent ($\alpha$) = 0.75, based
on the N$_2$-broadening parameters of CH$_3$CN \parencite{dudaryonok_15b}
and C$_3$H$_4$ \parencite{vinatier_07}. As these coefficients are not well known
for CH$_3$C$_3$N, we tested forward models over an appropriate parameter space 
[$\Gamma$ = 0.10--0.12 cm$^{-1}$
bar$^{-1}$; $\alpha$ = 0.65--0.85], but found these changes had little
effect (\textless 0.05$\%$) on the model reduced-$\chi^2$ values.

\begin{figure}
  \centering
  \includegraphics[scale=0.6]{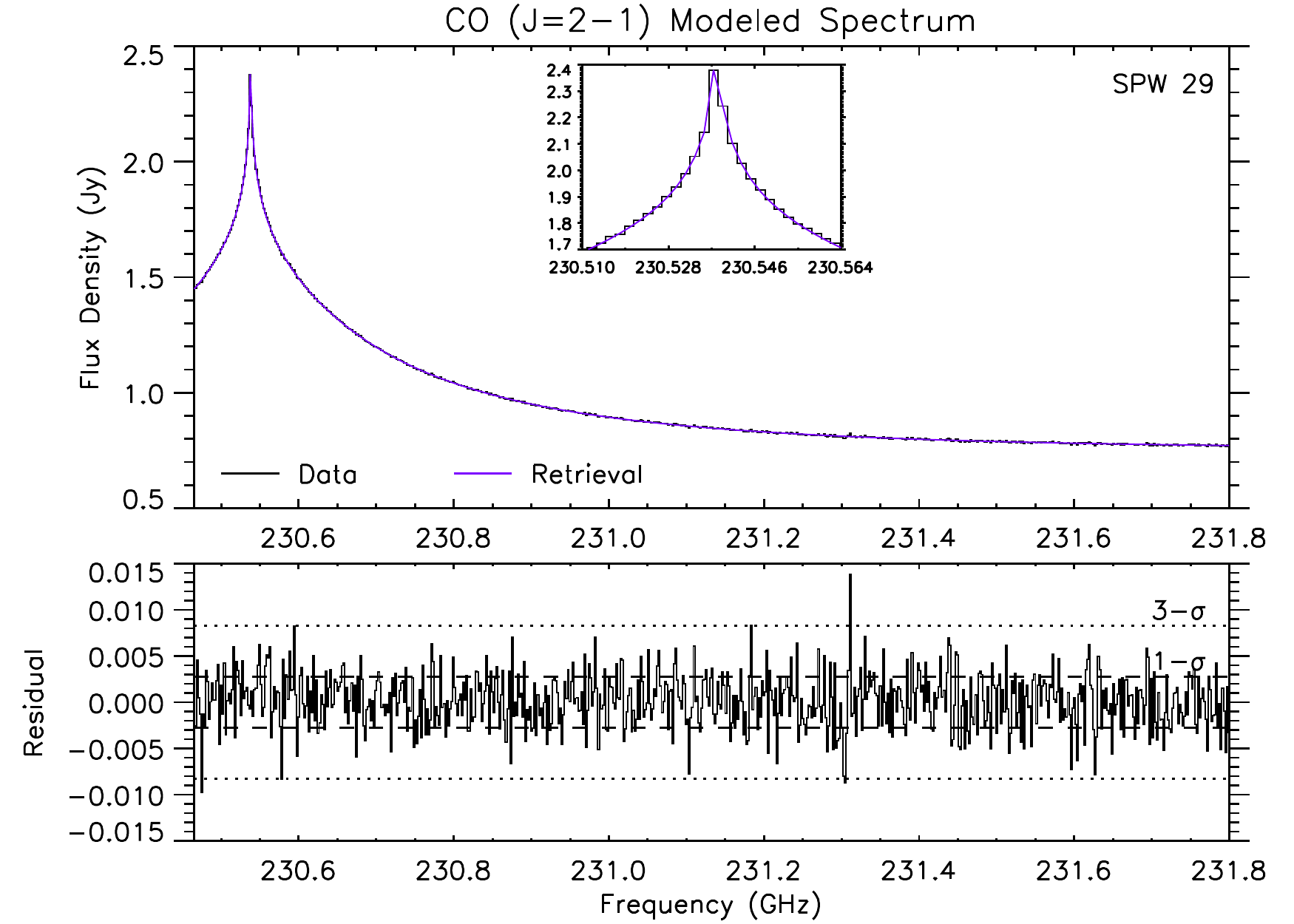}
  \includegraphics[scale=0.6]{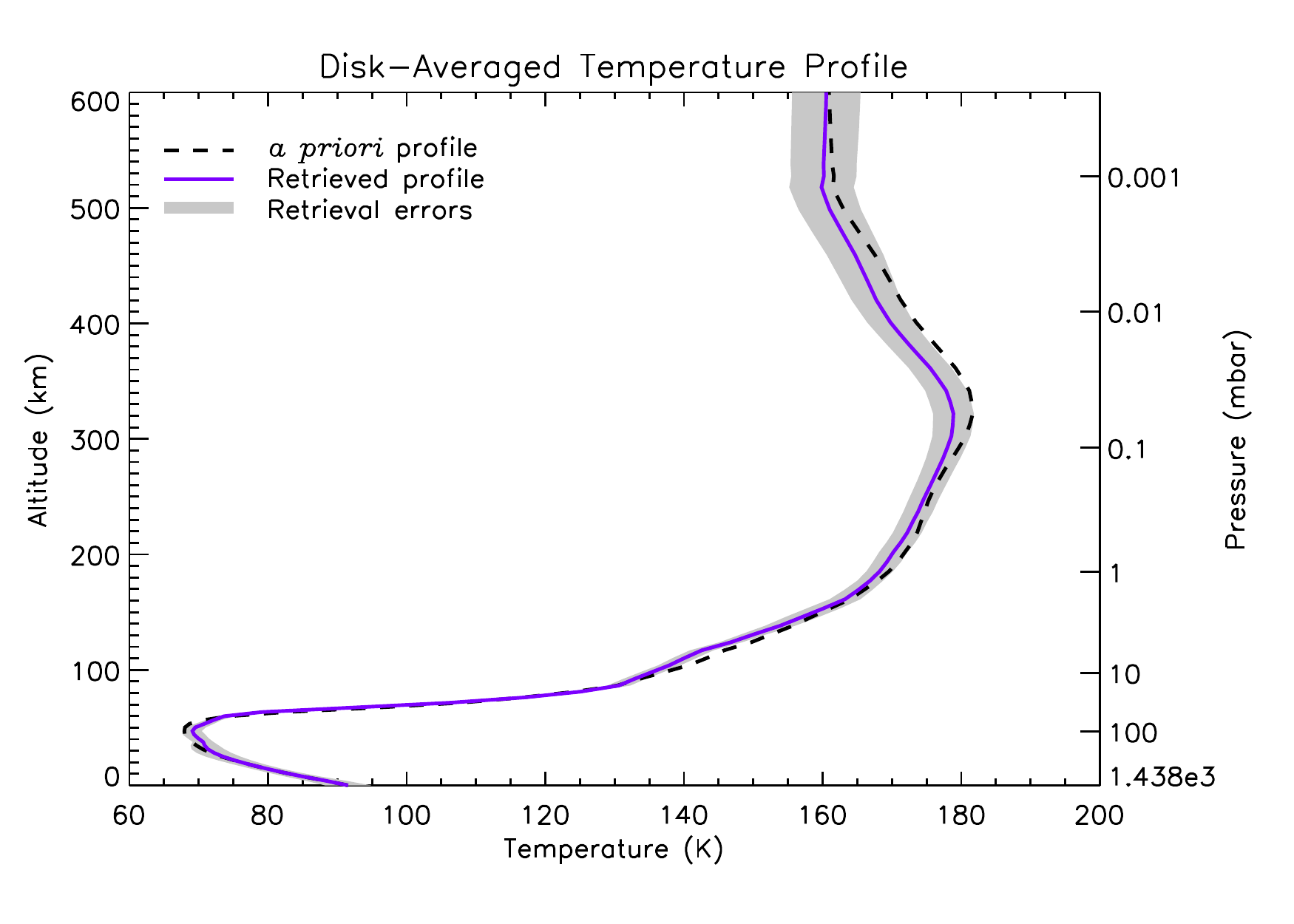}
  \caption{(Top) ALMA disk-averaged spectrum (black) of the CO $J = 2\rightarrow1$
    transition, with the best-fit NEMESIS model after retrieving
    Titan's vertical temperature profile (purple). The residual
    (data minus model) spectrum is shown below with 1$\sigma$ (dashed) and
    3$\sigma$ (dotted) RMS values. (Bottom) The corresponding
    retrieved temperature profile (purple) and error envelope
    (gray). The \textit{a priori} temperature profile is shown (dashed
  black) from previous ALMA and Cassini observations.}
  \label{fig:co_temps}
\end{figure}

We first retrieved Titan's disk-averaged
temperature profile by modeling the CO $J =
2\rightarrow1$ transition from SG1 SPW 29 (Fig. \ref{fig:co_temps},
top panel) by holding
the CO vertical volume mixing ratio constant at 49.6 ppm due to the
molecule's long photochemical lifetime in Titan's atmosphere \parencite{serigano_16,
  thelen_18}. The \textit{a priori} temperature profile was produced
through a combination of the retrieved disk-averaged profile from 
ALMA observations of Titan in 2015 \parencite{thelen_18} from lower
stratospheric altitudes through the mesosphere ($\sim$100--600 km), and from the 
Cassini radio-science and Huygens Atmospheric Structure Instrument
temperature measurements in the troposphere \parencite{fulchignoni_05,
  schinder_12}. The temperature profile was allowed to vary
continuously throughout the atmosphere, with \textit{a priori}
uncertainties initially set to 5
K and a correlation length of 1.5 scale heights to sufficiently reduce
artificial vertical oscillations in the retrieved
profile. The resulting temperature profile is shown in
Fig. \ref{fig:co_temps} (bottom panel), which was then used to model the CH$_3$C$_3$N
spectral bands from SG2 and SG3. As noted in previous ALMA studies,
the (sub)mm lines of CO in Titan's atmosphere allow for the
measurement of temperature throughout the stratosphere and into the
lower mesosphere, which is most notable in Fig. \ref{fig:co_temps}
(bottom panel)
where the retrieved temperature profile departs from the \textit{a
  priori} temperature profile at altitudes $\sim$100--530 km
($\sim$10--$10^{-3}$ mbar). Above 600 km, temperatures were set
as an isothermal profile at 160 K. 

\begin{figure}
\centering
\includegraphics[scale=0.5]{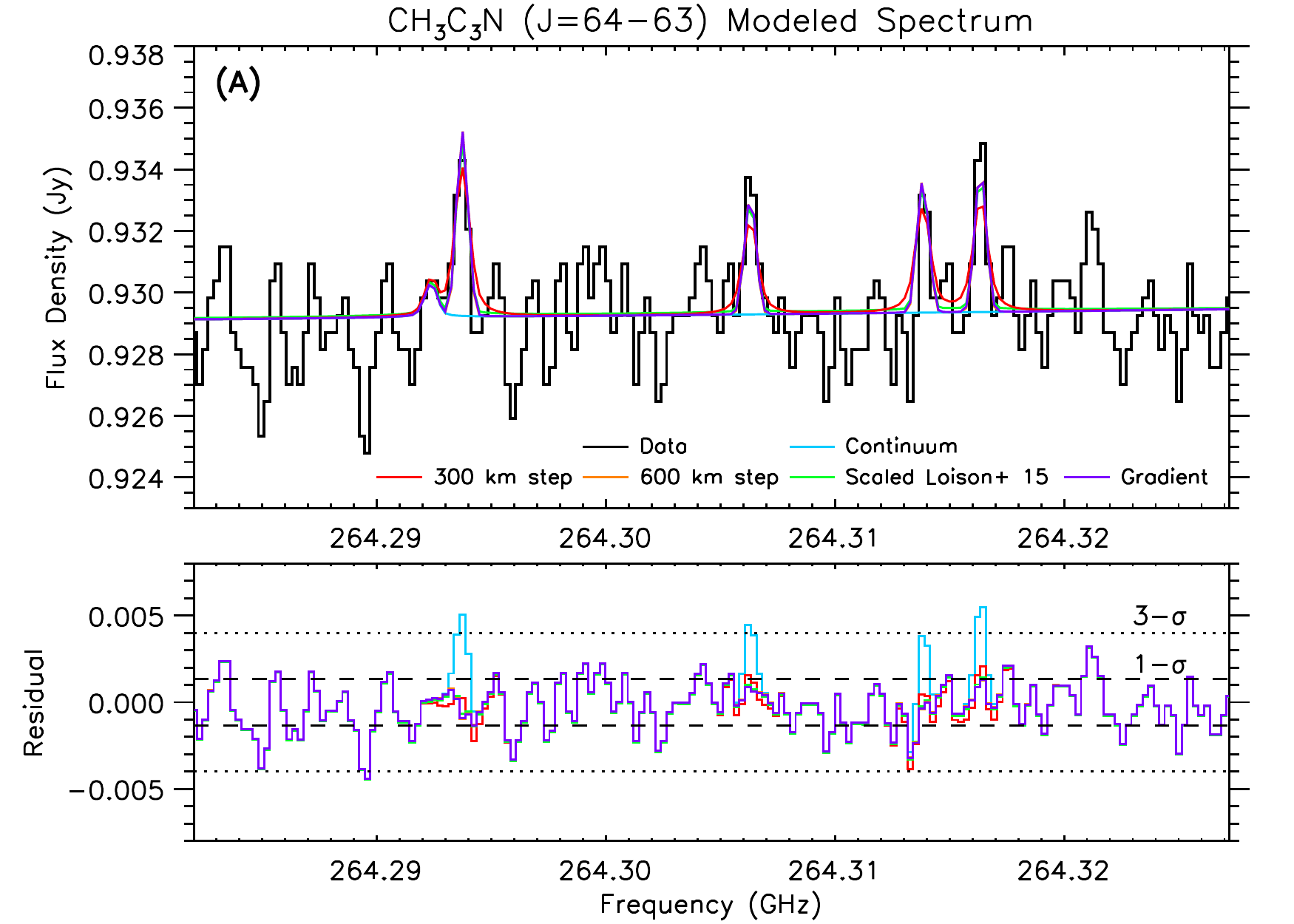}
\includegraphics[scale=0.5]{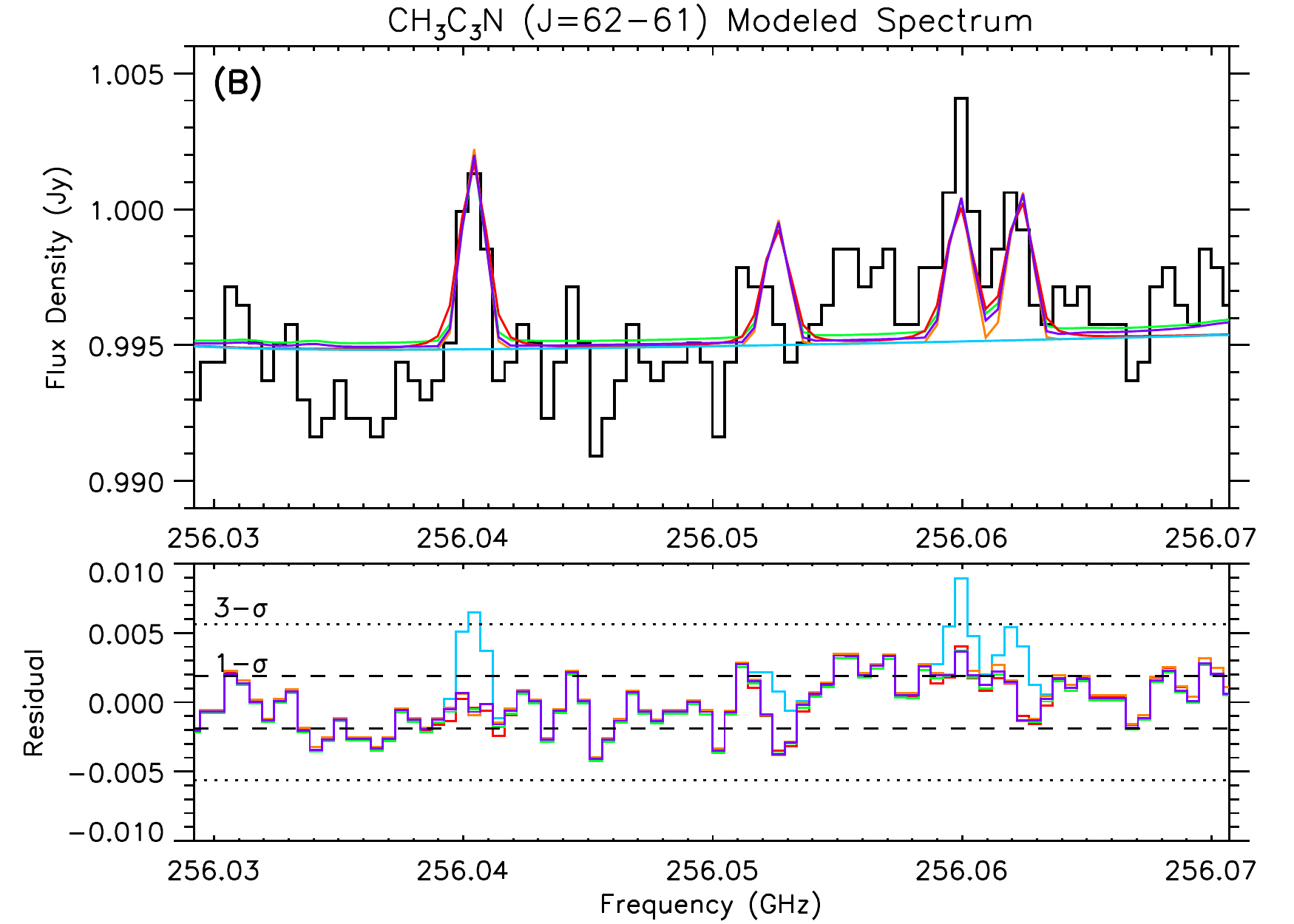}
\includegraphics[scale=0.5]{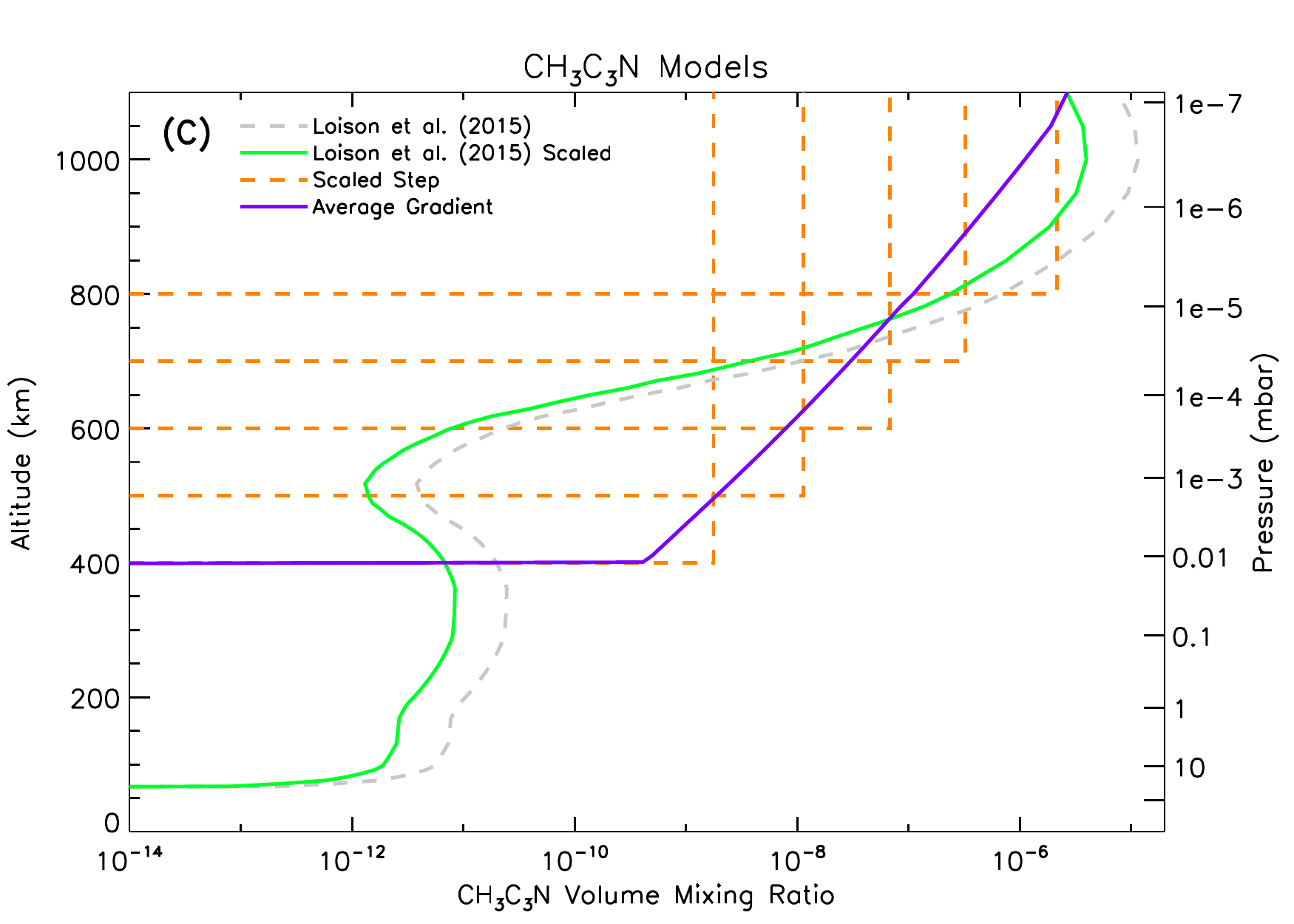}
\caption{(A) Disk-averaged spectrum of CH$_3$C$_3$N $J =
  64\rightarrow63$ (black) compared to various NEMESIS synthetic
  spectra (colored lines). The residual (data minus model) spectra are
  shown below with 1$\sigma$ (dashed) and
    3$\sigma$ (dotted) RMS values. A transition of C$_2$H$_5$CN is
    included in the model at $\sim$264.292 GHz. (B) The CH$_3$C$_3$N 
  $J = 62\rightarrow61$ band with modeled spectra, as in
  A. (C) The weighted mean best-fit vertical
  profiles of Titan's CH$_3$C$_3$N volume mixing ratio as retrieved with NEMESIS
  from spectra in A and B (green, orange, and purple lines). The
  nominal photochemical model of \textcite{loison_15} is shown for
  comparison (dashed gray).}
\label{fig:models}
\end{figure}

As transitions from numerous other trace species
are found in both SG2 and SG3 spectral windows containing
CH$_3$C$_3$N (Fig. \ref{fig:spec}), we included the disk-averaged
 volume mixing ratio profiles of C$_3$H$_4$, HC$_3$N (and its isotopes), and
C$_2$H$_5$CN from previous ALMA studies of Titan's atmosphere \parencite{thelen_19,
  cordiner_15, lai_17} in models of CH$_3$C$_3$N bands to correctly fit the
continuum and contributions from nearby line wings. To mitigate the
influence of these interloping species and best constrain the
retrieved CH$_3$C$_3$N mixing ratio profiles, we only modeled spectral regions
covering the $K=$ 0--3 transitions, as higher energy lines were not
detected in either CH$_3$C$_3$N band. Due to the unknown
nature (both in shape and relative abundance) of the vertical CH$_3$C$_3$N mixing ratio profile, we
attempted to fit both detected spectral bands with a variety of
vertical profiles. Previous ALMA studies
found that relatively narrow spectral lines
(such as C$_2$H$_3$CN, C$_2$H$_5$CN, c-C$_3$H$_2$) that sound Titan's
upper stratosphere and mesosphere could be adequately fit using vertical profiles
consisting of constant mixing ratios above a certain
altitude (step profiles), or profiles that are linear in log-pressure
space \parencite{cordiner_15,palmer_17,teanby_18,nixon_20}. Additionally,
photochemical models of Titan's atmosphere \parencite{loison_15,
  vuitton_19} make predictions for the vertical profile of CH$_3$C$_3$N and other trace
species, which can then be tested through radiative transfer
modeling. As the spectral resolution in these ALMA observations are
relatively high and include few spectral lines, we ran NEMESIS in
the more accurate line-by-line mode as opposed to utilizing the correlated-\textit{k}
method as is done for infrared and visible wavelengths. We fit
both spectral windows separately for independent
confirmation of the retrieved CH$_3$C$_3$N volume mixing ratio profiles, and in all
cases found the resulting mixing ratios
between the two spectral windows to agree within the 1$\sigma$
retrieval errors. We report the final volume mixing ratios as a weighted mean of
each pair of retrievals; the RMS of our SG2 data is $\sim\sqrt2$
less than that of SG3 (see Appendix
\ref{sec:app}). A variety of synthetic model spectra corresponding to the
vertical profile retrievals detailed below are shown for both bands of
CH$_3$C$_3$N in Fig. \ref{fig:models}A and B, with the weighted mean
best fit
profiles shown in Fig. \ref{fig:models}C compared to the photochemical
model of \textcite{loison_15}. The retrieved abundances and
calculated column densities for these profiles are detailed in Table 2.

First, we attempted to fit both CH$_3$C$_3$N spectral bands with a range of
step profiles from 100--800 km with uniform mixing ratio at every
100 km interval, initially set at a test value of 2.5 ppb. Profiles
were then scaled iteratively by NEMESIS until converging upon a fit
that sufficiently minimized the reduced-$\chi^2$ value, which was
found to be similar for all step profiles above 400 km (Table 2). Below these
altitudes, the synthetic CH$_3$C$_3$N line wings contribute too much to obtain a
good fit (i.e. $\chi^2/n>1$); an example is shown in
Fig. \ref{fig:models}A and B for a 300 km step model (red spectrum). Between 400--800 km, the spectral fits do not differ
significantly (Table 2). Here, we find the total
integrated column density of CH$_3$C$_3$N to range between
(3.86--5.73)$\times10^{12}$ cm$^{-2}$ from the best fit step models
(Fig. \ref{fig:models}C, orange dashed lines).

\begin{table}
  \begin{center}
    \caption[]{CH$_3$C$_3$N Best Fit Model Results}
    \begin{tabular}{lccc}
      \toprule
      \textbf{Model} &  \textbf{$\chi^2/n$$^{a}$}  &
                                                     \textbf{VMR$^{b}$} & \textbf{N (cm$^{-2}$)$^{c}$} \\
      \midrule
      \midrule
      400 km Step & 1.035 & $(1.771\pm0.196)\times10^{-09}$ & $5.727\times10^{12}$\\
      500 km Step & 1.020 & $(1.139\pm0.120)\times10^{-08}$ & $4.723\times10^{12}$\\
      600 km Step & 1.018 & $(6.851\pm0.693)\times10^{-08}$ & $4.400\times10^{12}$\\
      700 km Step & 1.018 & $(3.230\pm0.312)\times10^{-07}$ & $3.860\times10^{12}$\\
      800 km Step & 1.018 & $(2.152\pm0.198)\times10^{-06}$ & $5.607\times10^{12}$ \\
      \\
      Linear Gradient & 1.019 & & $4.709\times10^{12}$ \\
      \phantom{spce} Point 1 (400 km) & & $(4.093\pm0.339)\times10^{-10}$ &  \\
      \phantom{spce} Point 2 (1100 km) &  & $(2.675\pm0.222)\times10^{-06}$ & \\

      \\
      Loison et al. (2015) Scaling & 0.978 & $0.343\pm0.115$ & $9.741\times10^{12}$\\
      \bottomrule
    \end{tabular}
  \end{center}
  {\small{{\bf{Notes:}} $^a$ Reduced-$\chi^2$ values, where $n$ =
      number of data points minus model degrees of freedom. $^b$ Retrieved volume
      mixing ratios (VMR) are presented for all models except the scaling retrieval of the
      \textcite{loison_15} model, where the scale factor and error are
      shown. The VMR values for the two (high and low) pressure
      point fits of
      the linear gradient model are listed. $^c$ Total column density
      integrated up to 1200 km.}}
\end{table}

Next, a linear gradient was parameterized by allowing NEMESIS to vary
the volume mixing ratio and pressure between two points, with zero abundance below
the high-pressure point (Point 1, with pressure $p_1$ and mixing ratio
$q_1$) and constant abundance above the low-pressure point (Point 2,
with pressure $p_2$ and mixing
    ratio $q_2$). While the $p_1$ and $q_1$ values were initially
set with fairly arbitrary values with 
large errors to allow flexibility in the profile to achieve a good
fit, $p_2$ and $q_2$ were set to be constrained by the INMS measurements of
C$_4$H$_3$NH$^+$ ions in Titan's upper atmosphere ($\sim$1100 km),
with the inferred neutral C$_4$H$_3$N volume mixing ratio ranging between
(2--4)$\times10^{-6}$ \parencite{vuitton_07, vuitton_19}. Here, CH$_3$C$_3$N lines were only sensitive to
abundance above 400 km ($p_1 = 1.19\times10^{-2}$ mbar), resulting in $q_1 = 0.41$ ppb and $q_2 = 2.68$
ppm at 1100 km ($p_2 = 7.64\times10^{-8}$ mbar). The resulting
synthetic spectra and gradient profile are shown in
Fig. \ref{fig:models} (purple lines). The integrated column density of
this linear gradient model is 4.71$\times10^{12}$ cm$^{-2}$, in broad agreement
with the step model profiles.

Finally, we attempted to fit the spectra by retrieving a
multiplicative scaling factor applied to the photochemical model
profile of \textcite{loison_15}. This profile produced a good fit
when scaled by a factor of 0.34 of the original nominal model
(Fig. \ref{fig:models}A and B, green spectra; Fig. \ref{fig:models}C,
green line). The resulting
column density of $9.74\times10^{12}$ cm$^{-2}$, however, is 
higher than that of the best fit step or gradient models by a factor
of $\sim1.7$--2.5 (Table 2).

\section{Discussion $\&$ Conclusions} \label{sec:dis}
Though we were able to fit both detected CH$_3$C$_3$N spectral bands
with a variety of vertical profiles (Fig. \ref{fig:models}C), the
relatively short photochemical lifetime of CH$_3$C$_3$N -- between
$10^4$--$10^6$ seconds from 400--800 km \parencite{loison_15} -- suggests
that a vertically uniform mixing ratio profile may not be physically
realistic. As such, the scaled profile of \textcite{loison_15} and linear
gradient (Fig. \ref{fig:models}C, green and purple lines,
respectively) are favored for the volume mixing ratio profile of
CH$_3$C$_3$N. These profiles depict the formation of CH$_3$C$_3$N in
Titan's upper atmosphere 400--500 km (similar to its protonated counterpart,
C$_4$H$_3$NH$^+$) with decreasing abundance as a function of depth 
as the result of photodissociation and lack of reactive radicals (such
as CN and CCN). The dissociation of
CH$_3$C$_3$N has yet to be studied in detail, though the pathways
CH$_2$C$_3$N + H or CH$_3$ + C$_3$N have been suggested
by \textcite{loison_15}; alternatively, by
analogy with HC$_3$N (see \cite{huebner_15, vuitton_19}), we might expect
that CH$_3$C$_3$N photolysis instead yields CH$_3$C$_2$ and
CN. We find insufficient CH$_3$C$_3$N abundance at altitudes $<400$
km to properly identify the
dependence of the mixing ratio on altitude in Titan's
stratosphere and below, where GCR chemistry may
play an additional role in complex molecule formation.

Our retrieved volume mixing ratios above 700 km (Table 2) are in good agreement
with the estimated CH$_3$C$_3$N upper limit of $2.5\times10^{-7}$ by
\textcite{cerceau_85} based on the derived stratospheric HCN and HC$_3$N abundances in
Titan's north pole (then in winter) from \textit{Voyager-1} infrared
measurements. Further, the
derived column densities from this work between
(3.8--5.7)$\times10^{12}$ cm$^{-2}$ are in agreement with the lower value of
5.5$\times10^{12}$ cm$^{-2}$ found in the laboratory simulations by
\textcite{coll_99}. The scaling of the nominal \textcite{loison_15} profile by a
factor of 0.34 places it within 50$\%$ of their simulated profiles
(see their Fig. 14), which show significant spread due to the unknown
reaction rate coefficients for the production and loss of
CH$_3$C$_3$N. The linear gradient low-pressure point ($q_2$), 800 km step, and scaled profile of the
\textcite{loison_15} model results here are all in agreement with the inferred C$_4$H$_3$N
volume mixing ratio of $2\times10^{-6}$ at 1100 km from the Cassini T40 flyby
INMS measurements by \textcite{vuitton_19}. 

While the \textcite{loison_15} CH$_3$C$_3$N model corroborates the upper atmospheric
abundance of C$_4$H$_3$N inferred by \textcite{vuitton_07} from the T5
INMS measurements (a factor of 2 higher than those derived from T40 in
\cite{vuitton_19}), a large disparity
between the photochemical models (and within the ensemble of models produced by
\cite{loison_15}) arises in the lower atmosphere due to the poorly
constrained C$_4$H$_3$N branching ratios and reaction rate
coefficients at temperatures appropriate for Titan. Aside from electron dissociative
recombination of C$_4$H$_3$NH$^+$ \parencite{vuitton_07}, neutral production of
CH$_3$C$_3$N can occur in a few ways, as found through crossed beam
experiments, theoretical and
photochemical modeling studies \parencite{huang_99, balucani_00b, zhu_03, wang_06,
  loison_15}.
First, through the reactions of larger hydrocarbons with CN radicals:
\begin{eqnarray}
  \ce{CH3CCH + CN -> CH3C3N + H}, \label{eq:cn-a} \\
  \ce{CH3CCCH3 + CN -> CH3C3N + CH3}. \label{eq:cn-b}
\end{eqnarray}
Similarly, with CCN
radicals following their formation through H + HCCN \parencite{osamura_04,
  takayanagi_98} and subsequent
reactions with ethylene:
\begin{eqnarray}
  \ce{CCN + C2H4 -> CH3C3N + H}, \label{eq:c2n-a}
\end{eqnarray}
or through the chain beginning with acetylene:
\begin{eqnarray}
  \ce{CCN + C2H2 -> HC4N + H}, \nonumber \\
  \ce{HC4N + H ->[M] CH2C3N}, \nonumber \\
  \ce{CH2C3N + H ->[M] CH3C3N}. \label{eq:c2n-b}
\end{eqnarray}
While both reactions \ref{eq:c2n-a}
and \ref{eq:c2n-b} are found to be equally likely by \textcite{loison_15}, the
production of CCN via H + HCCN is not well constrained, and the
synthesis of CH$_3$C$_3$N through CN radicals
(Eqs. \ref{eq:cn-a},\ref{eq:cn-b}) are not included in their 
photochemical model. Additionally, cyanoallene may be produced
through reactions \ref{eq:cn-a}--\ref{eq:c2n-b} instead of (or in
addition to) methylcyanoacetylene. CH$_3$C$_3$N itself may form the protonated
species, C$_4$H$_3$NH$^+$, through reactions with the HCNH$^+$ and
C$_2$H$_5^+$ ions producing HCN and C$_2$H$_4$, respectively
\parencite{vuitton_07}. The other mechanism for forming C$_4$H$_3$NH$^+$
is through the combination of HCN and l-C$_3$H$_3^+$, though the
reaction rate
coefficient for this reaction and the abundance of l-C$_3$H$_3^+$ are
unknown \parencite{vuitton_07}. As such, the production and loss pathways
for both C$_4$H$_3$NH$^+$ and CH$_3$C$_3$N require further investigation.

The detection of CH$_3$C$_3$N here supports the previous
identification of C$_4$H$_3$NH$^+$ at $m/z=66$ from Cassini/INMS observations, and adds a 
valuable component to Titan's extensive atmospheric
photochemistry while revealing the need for further laboratory and
photochemical model studies detailing the production and dissociation of
Titan's larger nitriles. The retrieved column density and upper atmospheric
abundances agree with previous INMS measurements, laboratory and photochemical model
predictions, though the lack of sensitivity to Titan's lower
atmosphere through the $J = 64\rightarrow63$ and $J =
  62\rightarrow61$ rotational bands inhibits our investigation of the
 stratospheric volume mixing ratio and condensation of CH$_3$C$_3$N. However,
  these results provide insights into the possible shape of the full
  vertical profile through the scaling of the model produced by
  \textcite{loison_15}, and place constraints on the total column density
  of CH$_3$C$_3$N in Titan's atmosphere to aid in the determination of the production
   ratio of methylcyanoacetylene to cyanoallene, and the abundance of products
  resulting from the photodissociation
  both species. The detection of CH$_3$C$_3$N also provides the
  incentive for future observations of
  Titan at long wavelengths in the pursuit
of further complex, polar nitriles (such as C$_3$H$_7$CN and HC$_5$N) that
are predicted to exist in Titan's atmosphere. Finally, as with other trace
  species with fairly short photochemical lifetimes (compared to
  dynamical timescales), CH$_3$C$_3$N may have a complex and
  temporally variable spatial distribution that can be investigated
  with ALMA in the future through higher angular resolution observations.

  \section{Acknowledgments}
  The authors would like to thank N. Chanover, M. Dobrijevic, and
  J.C. Loison for their previous efforts in the proposed ALMA
  observations targeting CH$_3$C$_3$N and other molecules in Titan's
  atmosphere, and to R. Cosentino for discussions regarding ALMA
  bandpass smoothing.
  
AET was supported by the NASA Astrobiology Postdoctoral Program in
association with the NASA Astrobiology Institute (NAI) and the 
Universities Space Research Association. MAC was funded by the National
Science Foundation Grant $\#$AST-1613987. 
CAN and MAC received funding from NASA's Solar System Observations
Program. CAN was supported by the NAI. NAT and
PGJI were funded by the UK Science and Technology Facilities Council.

This paper makes use of the following ALMA data:
ADS/JAO.ALMA$\#$2019.1.00783.S. ALMA
is a partnership of ESO
(representing its member states), NSF (USA) and NINS (Japan), together
with NRC (Canada) and NSC and ASIAA (Taiwan) and KASI (Republic of
Korea), in cooperation with the Republic of Chile. The Joint ALMA
Observatory is operated by ESO, AUI/NRAO and NAOJ. The National Radio
Astronomy Observatory is a facility of the National Science Foundation
operated under cooperative agreement by Associated Universities, Inc.

\begin{appendix}
  \section{Application of ALMA Bandpass Smoothing} \label{sec:app}
Bandpass calibration is practiced in radio and (sub)millimeter observations
through the use of an off-source target to remove frequency (and sometimes time)
dependent fluctuations in visibility amplitudes and phases across
spectral windows, which often manifest
as continuum ripples or undulations in target spectra. Proper bandpass
calibration can increase the dynamic range of (sub)mm images, and
facilitates the measurement of weak (or broad) spectroscopic features.
Often, quasi-stellar objects with well characterized properties are observed for short
durations (typically 2--30 minutes) with ALMA to remove
visibility artifacts as a function of frequency, which improves
variations in amplitude and phase to $\lesssim0.1\%$ and $\lesssim0.3$ deg in ALMA
Band 3--6, respectively (though edge channels are still routinely removed
due to large amplitude changes). 

  \begin{figure}
    \centering
    \includegraphics[scale=0.8]{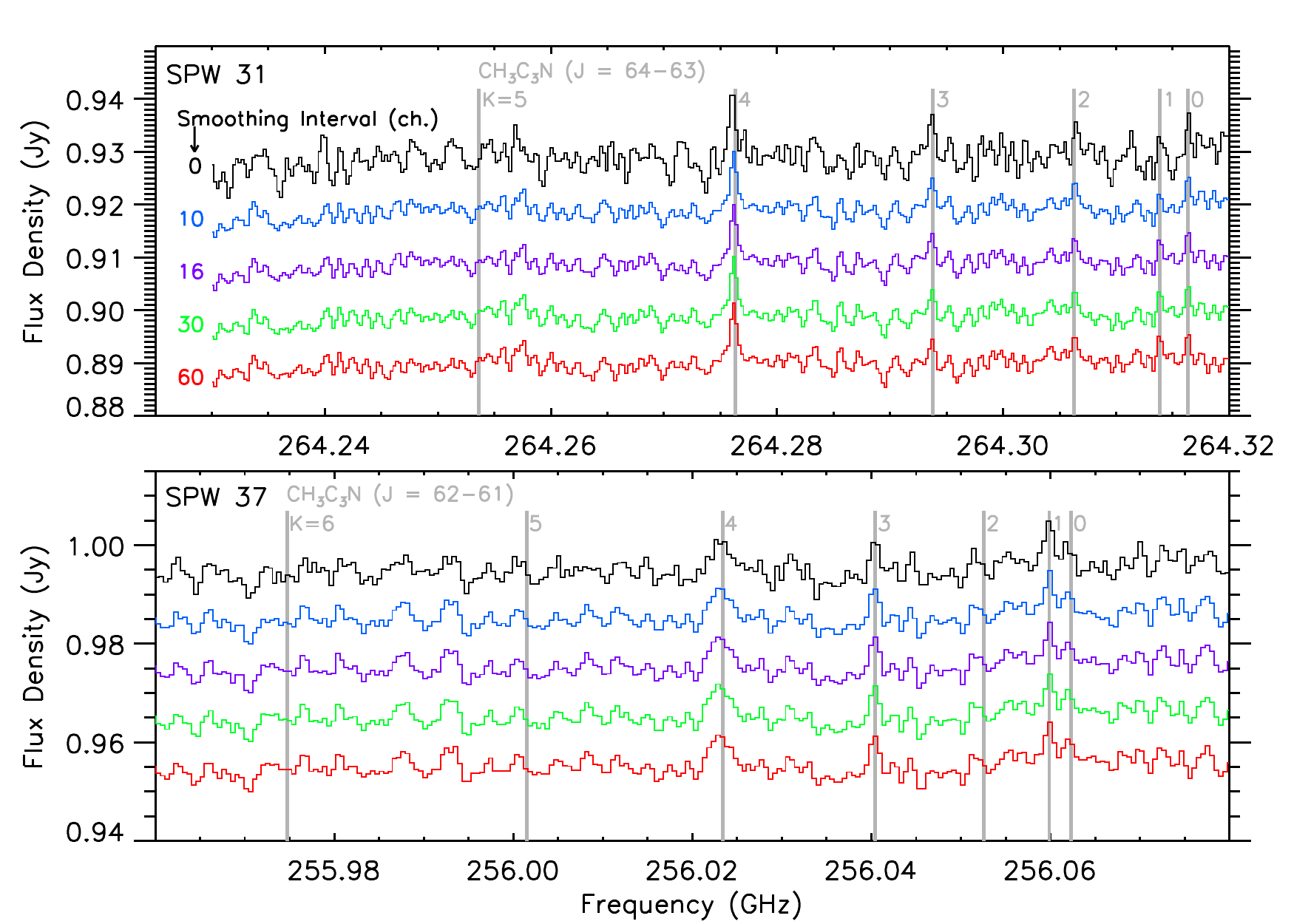}
    \caption{(Top) Disk-averaged Titan spectra from spectral window 31
    showing the effects of various bandpass smoothing
    intervals: 0 (black, default), 10 (blue), 16 (purple) 30 (green),
    and 60 channels (red). Spectra are separated by 10 mJy for
    visibility. Transitions of the CH$_3$C$_3$N $J =
      64\rightarrow63$ band are marked in gray. Spectral
      lines with significant flux (e.g. HC$_3$N $\nu_6=1$) were
      removed for clarity (see Fig. \ref{fig:spec}, Table 1). (Bottom)
      Spectra are shown as in the top panel, but for SPW 37. Smoothing was applied for the same
      number of channel intervals, and denoted by the same colors as
      the top panel. Transitions of CH$_3$C$_3$N $J =
      62\rightarrow61$ are marked in gray.}
    \label{fig:bpspec}
  \end{figure}

It has been shown that noise from variations in frequency in the bandpass calibration
solution can approach system noise (i.e. random noise in
radio antenna receivers) for short bandpass calibrator integrations,
but the application of 
bandpass smoothing (applied to the calibration target
solution) or additional
calibrator integration time can
further reduce the total spectral RMS through the reduction of
bandpass artifacts \parencite{yamaki_12}. This has been demonstrated to be
effective for ALMA as
well for spectral intervals with $\Delta\nu<100$ MHz\footnote{See ALMA
  Technical Notes 15: https://almascience.org/documents-and-tools/alma-technical-notes/ALMATechnicalNotes15$\_$FINAL.pdf/view}. As such,
though the total integration time of SG2 was a factor $\sim2$ more than that
of SG3 in our observations (see Section \ref{sec:obs}), the
corresponding spectral noise was not initially decreased by
$\sim$$\sqrt{2}$ due to the limitations of noise from the default bandpass
calibrator solutions. Here, we varied the bandpass
solution interval to apply smoothing to the bandpass
calibration function by averaging over a range of
channels. Fig. \ref{fig:bpspec} shows the resulting disk-averaged
spectra for SPW 31 and SPW 37 after using between 0--60 channel
bandpass smoothing solutions. Here, the effects of bandpass smoothing
are particularly evident in the comparison between 0 and
10--30 channel solution intervals.

  \begin{figure}
    \centering
    \includegraphics[scale=0.5]{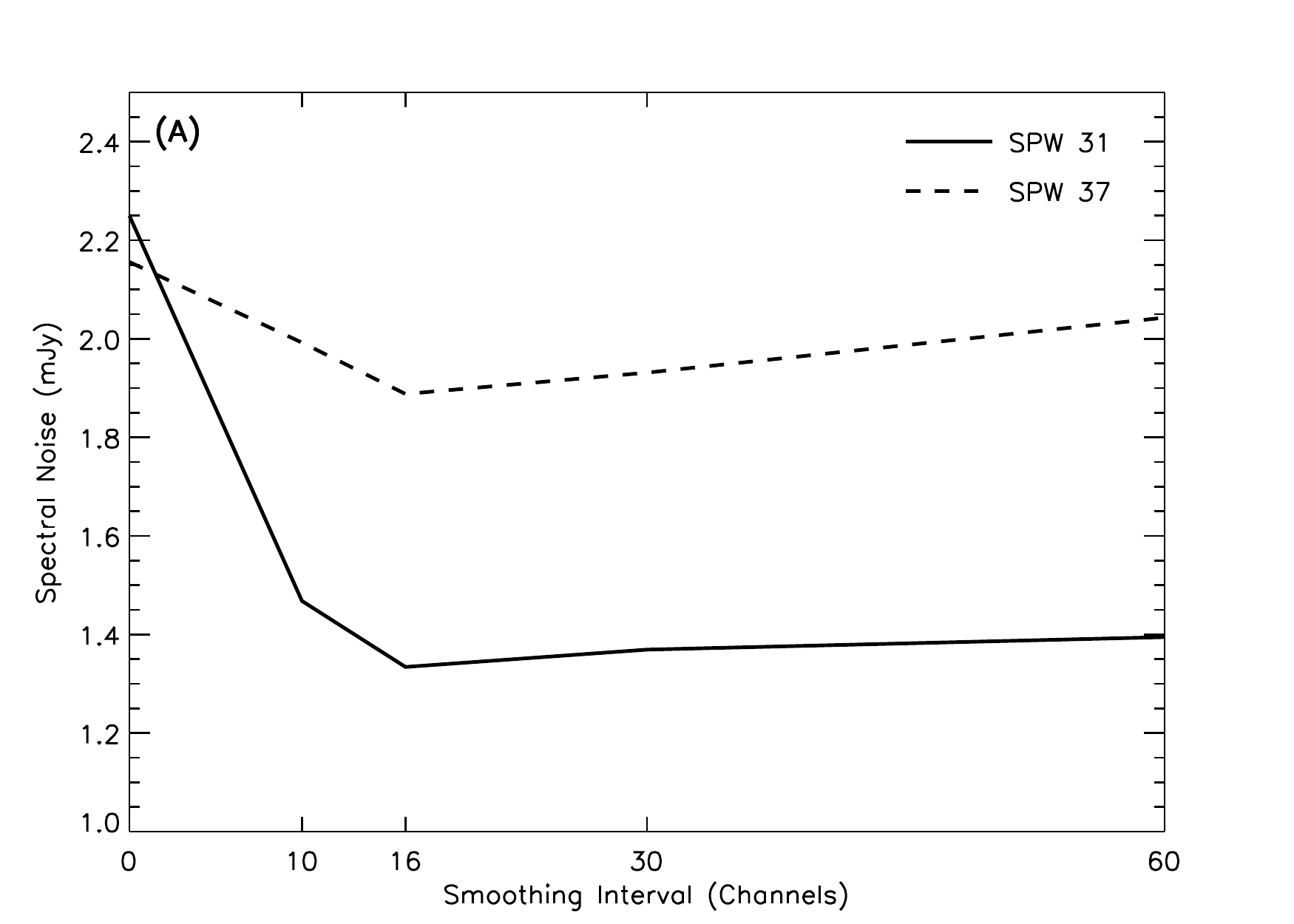}
    \includegraphics[scale=0.5]{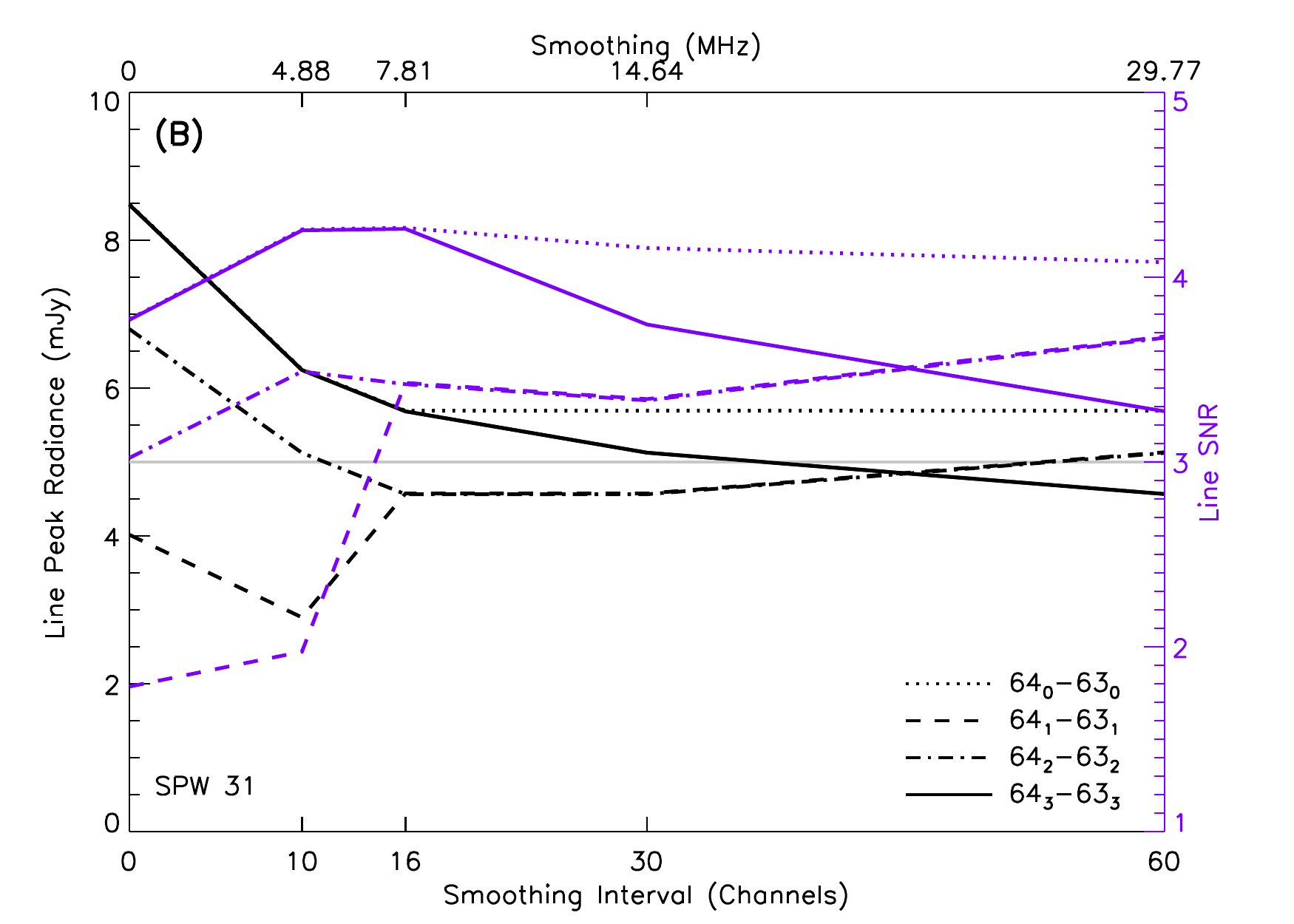}
    \includegraphics[scale=0.5]{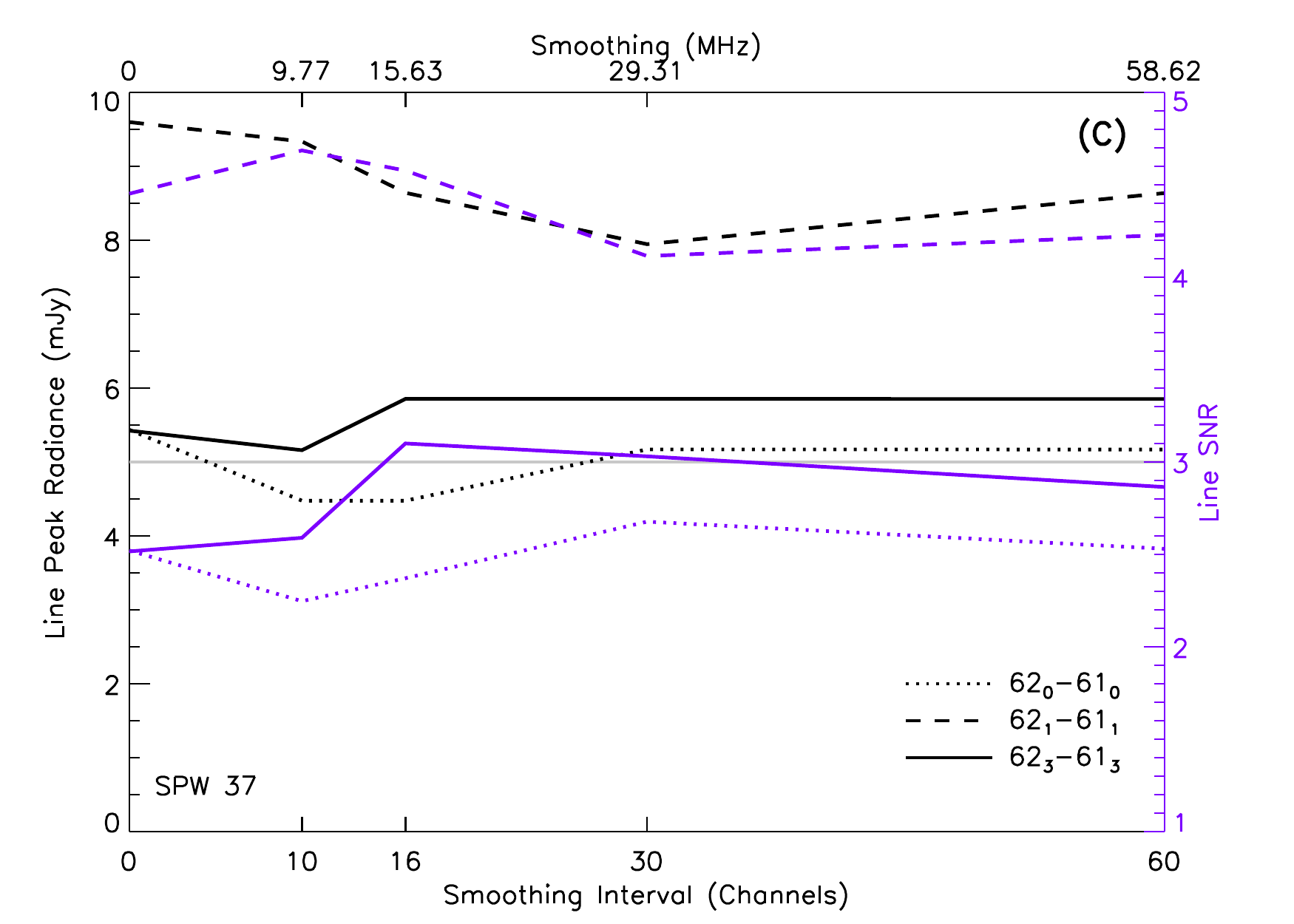}
    \caption{(A) The response of the root mean square
      noise in spectral window 31 (solid) and 37 (dashed) as a
      function of bandpass smoothing interval in channels. (B) The
      calculated line peak flux (line minus continuum) as a function of
      smoothing interval for all four detected
      CH$_3$C$_3$N transitions in SPW 31 (black lines) and the
      resulting signal-to-noise ratio (purple lines and corresponding
      y-axis) when divided by the RMS values
      in (A). The 3$\sigma$ threshold is marked in gray for reference. (C) Same as (B), but for SPW 37. The CH$_3$C$_3$N $J =
      62_2\rightarrow61_2$ transition is not plotted here, as the line
      flux remained at the noise level for all bandpass smoothing
      intervals (Figs. \ref{fig:models}B, \ref{fig:bpspec}).}
    \label{fig:bpsnr}
  \end{figure}

\begin{figure}
    \centering
    \includegraphics[scale=0.9]{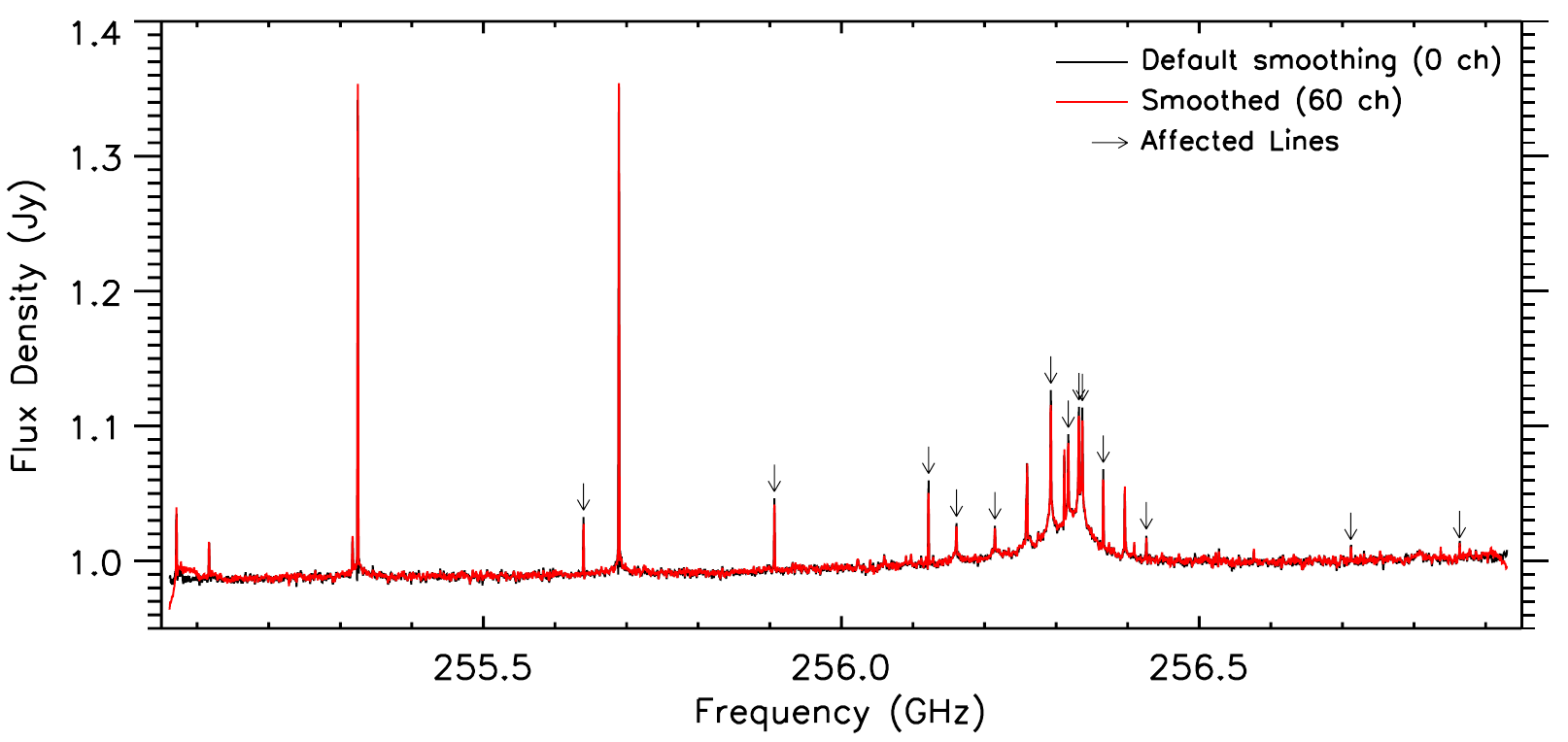}
    \caption{Comparison of disk-averaged Titan spectra from SG3 (SPW 37) with
      default (black) and 60 channel bandpass smoothing (red)
      applied. Transitions with reduced apparent line peak
      fluxes after the decrease of RMS noise as a result of bandpass
      smoothing are marked with arrows. Bandpass artifacts induced by
      excessive smoothing are apparent towards the bandwidth edges
      (red spectrum $<255.15$ GHz and $>256.9$ GHz).}
    \label{fig:bplines}
  \end{figure}

We found that after applying a smoothing interval of 16 channels
(7.81 MHz in SG2 SPW 31, 15.6 MHz in SG3 SPW 37) the RMS decreased by
a factor of 1.67 in SPW 31 and by a factor of
1.12 in SPW 37 (Fig. \ref{fig:bpsnr}A). The resulting RMS in
SPW 31 (1.33 mJy) was then a factor of $\sim$$\sqrt{2}$ less than in SPW
37 (1.88 mJy). Additionally, the decrease in spectral noise from the CH$_3$C$_3$N
bands reduced the apparent peak line flux density of some transitions by
$\sim1\sigma$, but the
corresponding decrease in RMS improved the overall SNR
in most lines of both spectral bands (Fig. \ref{fig:bpsnr}B,C). An
example of the removal of additive noise peaks from spectral line
fluxes after bandpass smoothing is
shown in Fig. \ref{fig:bplines}. We
found that 16 channel smoothing resulted in the optimal increase in
SNR across all lines of both spectral bands for
CH$_3$C$_3$N. Increased smoothing ($>20$--30 MHz) showed
diminishing returns on spectral RMS, though caution should be taken
when averaging over large intervals, as the continuum may be aversely
affected -- particularly for channels close to either edge of the
bandwidth (Fig. \ref{fig:bplines}, red spectrum). The optimal channel interval depends on spectral
resolution, frequency, bandpass and target integration time, so we
encourage observers to experiment with
multiple bandpass smoothing solutions to determine the appropriate
solution interval for observations with ALMA.

The application of bandpass smoothing in radio spectra has previously
been studied by \textcite{yamaki_12} with similar results in RMS
improvements after increased channel smoothing intervals and
additional time on bandpass calibration sources. To facilitate the
detection of additional trace gases in planetary atmospheres, bandpass
smoothing may be applied to ALMA data with long integration
times. Here, we find limited increase in line SNR for SG3
(Fig. \ref{fig:bpsnr}A and C), similar to
previous efforts to detect c-C$_3$H$_2$ in ALMA observations of Titan
\parencite{nixon_20}; however, the significant decrease in RMS in SG2
(Fig. \ref{fig:bpsnr}A and B)
promotes the use of bandpass smoothing for Titan observations with
total (on source) integration times of $\gtrsim2$ hours. Additionally,
increased integration time on bandpass
calibrators, which is available under specific well-justified conditions by ALMA, may further decrease the spectral RMS
\parencite{yamaki_12}.
  
\end{appendix}

\pagebreak
\printbibliography[title={References}]

\end{document}